\pdfoutput=1
\documentclass[prb,amssymb,twocolumn,superscriptaddress]{revtex4-1}

\usepackage{graphics}
\usepackage{mathtools}
\usepackage{float}
\usepackage{amsmath}

\begin{document}
\title{Magnetoresistance in the superconducting state at the (111) LaAlO$_3$/SrTiO$_3$ interface}

\author{S. Davis}
\email[]{samueldavis2016@u.northwestern.edu}

\affiliation{Graduate Program in Applied Physics and Department of Physics and Astronomy, Northwestern University, 2145 Sheridan Road, Evanston, IL 60208, USA}
\author{Z. Huang}
\author{K. Han}

\affiliation{NUSNNI-Nanocore, National University of Singapore 117411, Singapore}  
\affiliation{Department of Physics, National University of Singapore 117551, Singapore } 
\author{Ariando}
\affiliation{NUSNNI-Nanocore, National University of Singapore 117411, Singapore}  
\affiliation{Department of Physics, National University of Singapore 117551, Singapore } 
\affiliation{NUS Graduate School for Integrative Sciences \& Engineering, National University of Singapore 117456, Singapore}

\author{T. Venkatesan}
\affiliation{NUS Graduate School for Integrative Sciences \& Engineering, National University of Singapore 117456, Singapore}
\affiliation{Department of Electrical and Computer Engineering, National University of Singapore 117576, Singapore}\affiliation{Department of Material Science and Engineering, National University of Singapore 117575, Singapore}

\author{V. Chandrasekhar}
\email[]{v-chandrasekhar@northwestern.edu}
\affiliation{Graduate Program in Applied Physics and Department of Physics and Astronomy, Northwestern University, 2145 Sheridan Road, Evanston, IL 60208, USA}

\begin{abstract}
Condensed matter systems that simultaneously exhibit superconductivity and ferromagnetism are rare due the antagonistic relationship between conventional spin-singlet superconductivity and ferromagnetic order. In materials in which superconductivity and magnetic order is known to coexist (such as some heavy-fermion materials), the superconductivity is thought to be of an unconventional nature.  Recently, the conducting gas that lives at the interface between the perovskite band insulators LaAlO$_3$ (LAO) and SrTiO$_3$ (STO) has also been shown to host both superconductivity and magnetism. Most previous research has focused on LAO/STO samples in which the interface is in the (001) crystal plane.  Relatively little work has focused on the (111) crystal orientation, which has hexagonal symmetry at the interface, and has been predicted to have potentially interesting topological properties, including unconventional superconducting pairing states.  Here we report measurements of the magnetoresistance of (111) LAO/STO heterostructures at temperatures at which they are also superconducting. As with the (001) structures, the magnetoresistance is hysteretic, indicating the coexistence of magnetism and superconductivity, but in addition, we find that this magnetoresistance is anisotropic. Such an anisotropic response is completely unexpected in the superconducting state, and suggests that (111) LAO/STO heterostructures may support unconventional superconductivity.
\end{abstract}

\date{\today}%
\maketitle

Ferromagnetism is traditionally considered antithetical to conventional superconductivity: electron spin moments are aligned in a ferromagnet, while conventional $s$-wave superconductivity requires the anti-alignment of the spin moments of the two electrons in a Cooper pair.  While there do exist systems in which superconductivity and magnetism coexist, the pairing in these systems is thought to be of an unconventional nature: examples are heavy-fermion materials like UGe$_2$ and UPt$_3$ that are thought to have $p$ or $f$-wave orbital pairing in triplet spin states in some phases.\cite{Kote,Gannon}  More recently, ferromagnetism and superconductivity have been found to coexist at the LAO/STO interface.\cite{Dikin,Schne,Li,Bert}  Low temperature magnetotransport measurements show that the influence of the ferromagnetism on the superconducting state at the LAO/STO occurs primarily through the external magnetic field arising from the ferromagnet,\cite{Mehta} although the observation of anisotropic magnetoresistance (AMR) suggests that there is direct exchange coupling between the charge carriers and the magnetic moments at the interface.\cite{Joshua, Miao, Annadi}.

Most experiments on LAO/STO interfaces have been performed on samples where the interface is in the (001) crystal plane.\cite{Dikin,Schne,Li, Bert,Mehta,Joshua}  In the past few years, interest has grown in LAO/STO interfaces in the (111) crystal orientation.  The Ti atoms at the (111) LAO/STO interface have hexagonal symmetry,\cite{Rodel,Walker} and consequently have been predicted to have potential topological properties,\cite{Doe} and to possibly host time-reversal symmetry breaking superconducting pairing states.\cite{Scheu}  Electrical transport measurements on the conducting carrier gas at the (111) LAO/STO interface have found not only a superconducting state\cite{mont,davis3,rout2} but also the presence of strong anisotropy in almost all properties when measured along different surface crystal directions, including longitudinal resistance, Hall effect, quantum capacitance, AMR and superconductivity.\cite{davis2,davis1,rout,davis3}  Here we show that the magnetoresistance at or in the superconducting state is hysteretic and also anisotropic.  As with the (001) LAO/STO devices, the presence of a hysteretic magnetoresistance (MR) in the (111) LAO/STO devices indicates magnetism coexisting with superconductivity.\cite{Dikin,Mehta}  The anisotropic nature of the MR reinforces observation of the anisotropic superconducting properties seen previously, and raises the possibility that the superconducting state may be unconventional in nature.

\begin{figure}[!]
\center{\includegraphics[width=6cm]{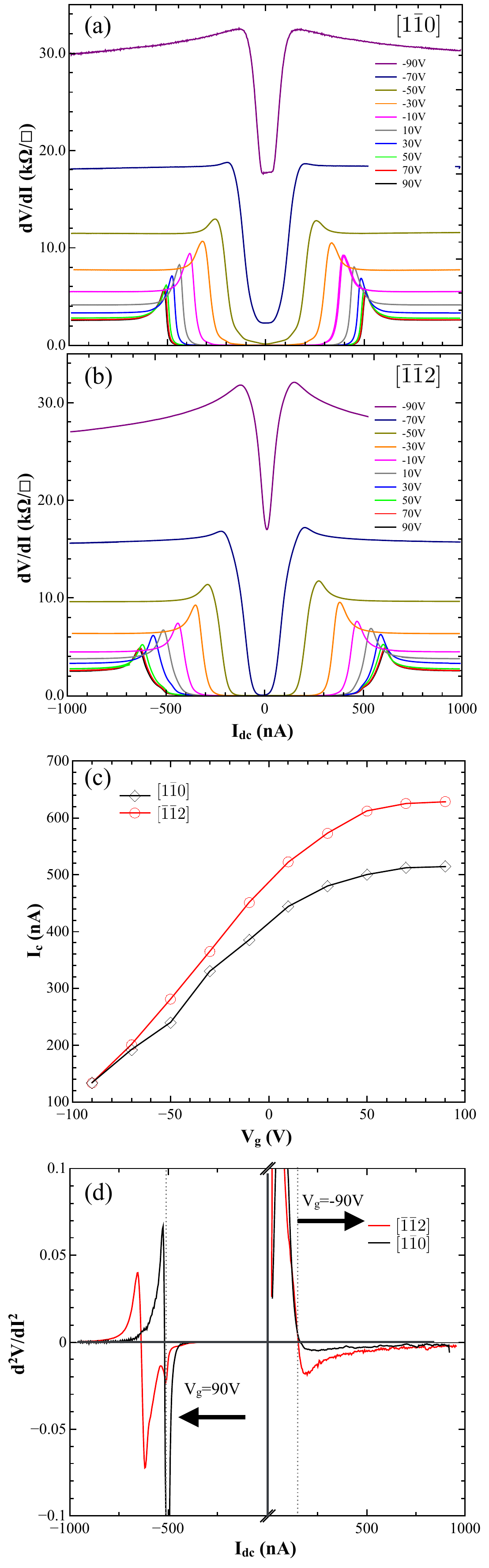}}
\caption{Longitudinal differential resistance $dV/dI$ vs dc current $I_{dc}$ at 30 mK at different gate voltages $V_g$ for Hall bars aligned along the [$1\bar{1}0$] crystal direction (a) and the [$\bar{1}\bar{1}2$] crystal direction (b).  (c)  Critical current $I_c$ for the two directions as a function of $V_g$.  Inset:  $R(T)$ along the two directions, showing the superconducting transition. (d) Second derivative $d^2V/dI^2$ vs. $I_{dc}$ along the two crystal directions for $V_g$=90 V (negative $x$ axis) and $V_g$=-90 V (positive $x$ axis). }
\label{dVdI} 
\end{figure}
The samples in this study consisted of four Hall bar devices fabricated on a single substrate, such that two Hall bars lie along the [$1\bar{1}0$] direction and two lie along the [$\bar{1}\bar{1}2$] surface crystal direction of the (111) LAO/STO interface.  Details about the film synthesis and sample fabrication can be found in earlier publications. \cite{davis2,davis1} The properties of the devices depend on post-growth treatment.  Three different post-growth treatments were discussed in an earlier publication:\cite{davis1}  annealing in a O$_2$ environment, annealing in a Ar/H$_2$ environment, and ultraviolet (UV) irradiation under ambient conditions.  It is only for the Ar/H$_2$ annealed samples that the devices show a zero resistance state along both surface crystal directions, and hence we discuss here only these devices.\cite{davis3}  We note, however, that the Ar/H$_2$ annealed devices have a far lower sheet resistance and show far less anisotropy than either as-grown samples or samples subjected to a O$_2$ anneal or UV irradiation;\cite{davis1} nevertheless, as we show below, the anisotropy is quite evident.  The Hall bars were measured in an Oxford Kelvinox MX100 dilution refrigerator using conventional lock-in techniques.  Both sets of devices showed quantitatively similar behavior; here we only report the results for two devices on which we performed the most extensive measurements.

Figures \ref{dVdI}(a) and (b) show the differential resistance $dV/dI$ as a function of dc current $I_{dc}$ at 30 mK for two Hall bars, one aligned along the [$1\bar{1}0$] direction, and one along the [$\bar{1}\bar{1}2$] direction, for various values of the back gate voltage $V_g$.  The devices were cooled from 4 K with a gate voltage $V_g=-10$ V applied.  As can be seen, both devices are superconducting for positive gate voltages, leaving the zero resistance state for large negative $V_g$ ($V_g\sim-50$V for the [$1\bar{1}0$] Hall bar, and $V_g\sim -70$V for the [$\bar{1}\bar{1}2$] Hall bar).  However, both Hall bars show a decrease in resistance near zero bias even at the largest negative $V_g = -90$ V, indicating the presence of superconducting correlations.  It is clear from these traces that the superconducting characteristics are anisotropic, even in these Ar/H$_2$ annealed samples which show much less anisotropy than O$_2$ annealed or UV irradiated devices.\cite{davis1}  The anisotropy is visible in the critical current $I_c$ (defined as the current $I_{dc}$ at which the maxima in $dV/dI$ occur in Figs. \ref{dVdI}(a) and (b)) as a function of $V_g$ for both Hall bars, which is shown in Fig. \ref{dVdI}(c).  While $I_c$ is the same for both directions at large negative $V_g$, $I_c$ becomes progressively larger for the [$\bar{1}\bar{1}2$] Hall bar in comparison to the [$1\bar{1}0$] Hall bar as $V_g$ is increased above -50 V.  Surprisingly, the superconducting transitions from the zero resistance state with increasing $|I_{dc}|$ are sharper for the [$1\bar{1}0$] direction than for the [$\bar{1}\bar{1}2$] direction for positive $V_g$, even though the [$1\bar{1}0$] direction has higher normal state resistance at all gate voltages.  Closer inspection of the $dV/dI$ vs. $I_{dc}$ traces for the [$\bar{1}\bar{1}2$] Hall bar show that for large positive values of $V_g$, a small peak appears at a value of $I_{dc}$ corresponding to $I_c$ in the [$1\bar{1}0$] direction.  This is more clearly seen by examining the second derivative $d^2V/dI^2$ vs. $I_{dc}$, shown in Figure \ref{dVdI}(d) for $V_g$=-90 and $V_g$=90 V.  $d^2V/dI^2$ is zero when $dV/dI$ is maximum, and hence the zero crossings of $d^2V/dI^2$ determine $I_c$.  For $V_g$=-90 V, $d^2V/dI^2$ crosses zero at approximately the same current for both crystal directions, and hence the $I_c$'s are also approximately the same.  For $V_g$=90 V, the zero crossings for the two crystal directions are quite different, corresponding to the different $I_c$'s seen in Fig. \ref{dVdI}(c).  However, $d^2V/dI^2$ vs $I_{dc}$ for the [$\bar{1}\bar{1}2$] direction shows a dip near the critical current for the [$1\bar{1}0$] direction, a signature of the small peak in the [$\bar{1}\bar{1}2$] $dV/dI$ traces mentioned above. 
This suggests that measurements along the [$\bar{1}\bar{1}2$] Hall bar for positive $V_g$ also sample the superconducting characteristics along the [$1\bar{1}0$] direction, although the reverse is apparently not true.  This may be the reason behind the broader transitions in the [$\bar{1}\bar{1}2$] Hall bar for large positive $V_g$.  

\begin{figure}[h!]
\includegraphics[width=7cm]{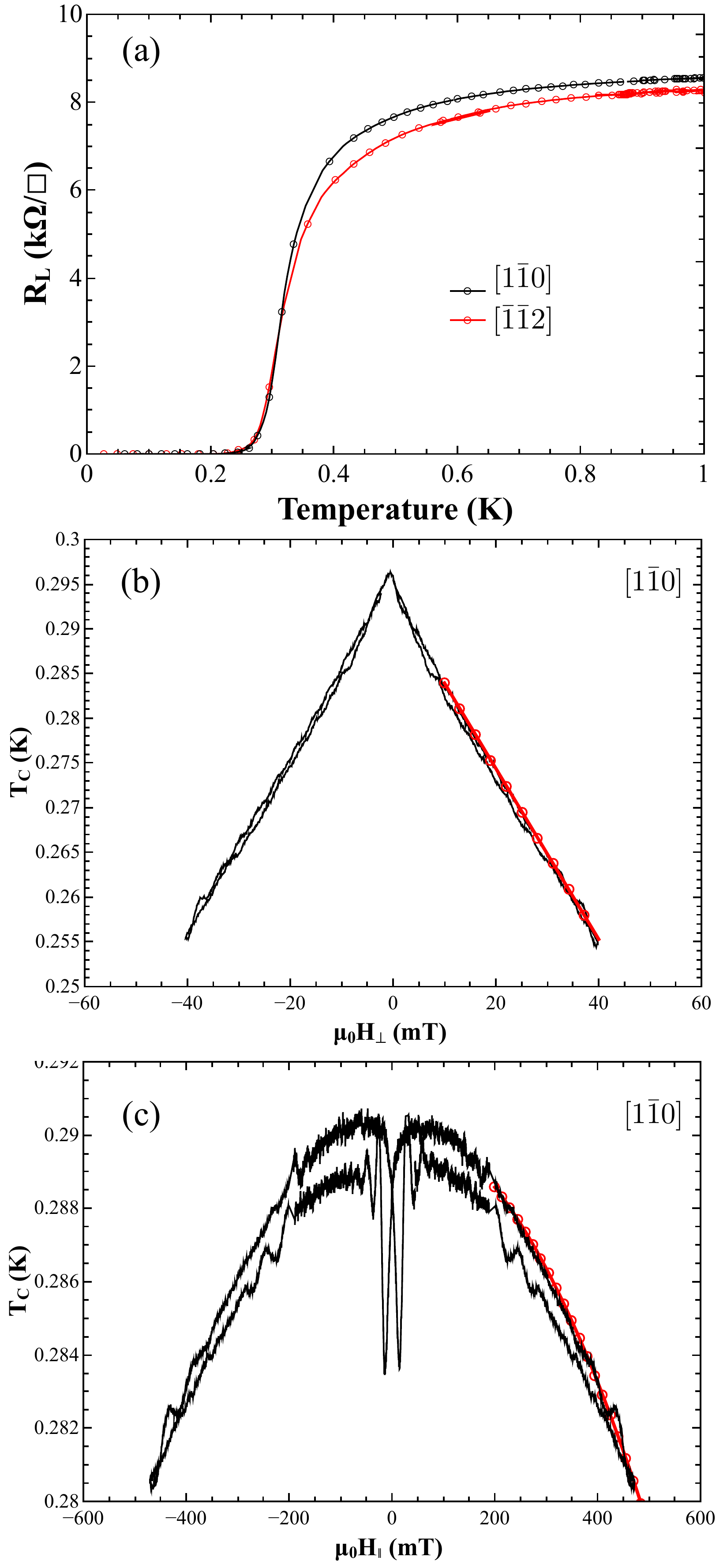}
\caption{(a)  Sheet resistance as a function of temperature for the [$1\bar{1}0$] and [$\bar{1}\bar{1}2$] Hall bars, with $V_g$=-10 V.  (b)  Superconducting transition temperature $T_c$ as a function of perpendicular magnetic field for the [$1\bar{1}0$] Hall bar for $V_g$=100 V.  The symboled line is a fit to the linear dependence.(c)  $T_c$ as function of parallel magnetic field for the [$1\bar{1}0$] Hall bar for $V_g$=100 V.  The symboled line is a fit to a quadratic dependence.}
\label{RofT}
\end{figure} 
The difference between the two crystal directions is also reflected in the temperature-dependent superconducting transition, shown in Fig. \ref{RofT}(a).   The transition in the [$\bar{1}\bar{1}2$] direction is broader, although the zero-resistance state is reached at the same temperature in both directions.  One can also obtain the Ginzburg-Landau superconducting coherence length $\xi_S$ and the thickness $d$ of the conducting gas by measuring phase diagram (superconducting transition temperature $T_c$ as a function of magnetic field) in parallel and perpendicular magnetic fields. \cite{Dikin,Mehta}  In a perpendicular field, the critical field $H_{c\perp}$ is roughly the field required to put one superconducting flux quantum $\Phi_0=h/2e$ in an area the size of $\xi_S^2$ /cite{Tinkham}
\begin{equation}
H_{c\perp} = \frac{\Phi_0}{2 \pi \xi_S^2(T)}
\end{equation}  
where the temperature dependent superconducting phase coherence length is given by
\begin{equation}
\xi_S(T) = \alpha \frac{\xi_0}{\sqrt{(1- T/T_c)}}.
\end{equation}
Here $\xi_0$ is the zero temperature coherence length, and $\alpha$=0.74 or 0.86 for the clean and dirty limit respectively.\cite{Tinkham}  The dependence of $H_{c\perp}$ on $T$ is therefore linear, with a slope given by
\begin{equation}
\frac{dH_{c\perp}}{dT} = - \frac{\Phi_0}{2 \pi \xi_0^2 T_c},
\end{equation} 
where we have taken $\alpha\sim1$.  Thus by measuring the slope of the phase diagram and $T_c$, one can determine $\xi_0$.  In a parallel field, for a superconductor whose thickness $d$ is less than $\xi_S$, the area is restricted by $d$, hence the parallel critical field is given by \cite{Tinkham}
\begin{equation}
H_{c\parallel} = \frac{\sqrt{3}\Phi_0}{\pi \xi_S(T) d},
\end{equation}
so that $T_c(H)$ should have a quadratic dependence on $H$.  Fitting this dependence, using the value of $\xi_0$ obtained from the perpendicular field phase diagram, one can obtain an estimate for the thickness $d$ of the superconductor.

Figures \ref{RofT}(b) and (c) show the phase diagrams for the Hall bar oriented along the [$1\bar{1}0$] crystal directions in magnetic fields perpendicular and parallel to the interface, for $V_g$=100 V (results along the [$\bar{1}\bar{1}2$] direction are similar).  These data were taken by maintaining the resistance of the device at the midpoint of the resistive superconducting transition while continuously sweeping the magnetic field.  Each of these traces took approximately 20 hours to complete.  As expected, the slope of $T_c$ vs $H_\perp$ is linear, and from the slope, we obtain a zero temperature coherence length of $\xi_0$= 26 nm.  The phase diagram in parallel field is more complicated:  while the overall background is quadratic as expected, near zero field, we observe hysteresis in the field dependence, with sharp dips in $T_c$ near zero field.  This hysteresis is a result of the hysteresis in the MR, which is discussed in detail below, and is similar to what is observed in (001) LAO/STO devices.\cite{Dikin,Mehta}  Fitting the data for $|\mu_0 H_\parallel |>$ 50 mT to the expected quadratic form, we obtain a superconducting film thickness of $d \sim$7 nm.  This value is for $V_g$=100 V, where the conducting gas thickness is expected to be maximum, so that the film thickness at other gate voltages will be smaller.  We note that this value is smaller than the values typically reported in (001) LAO/STO, which usually are of order 10-15 nm.\cite{mont,Dikin,Mehta}      

\begin{figure}[h!]
\includegraphics[width=6.5cm]{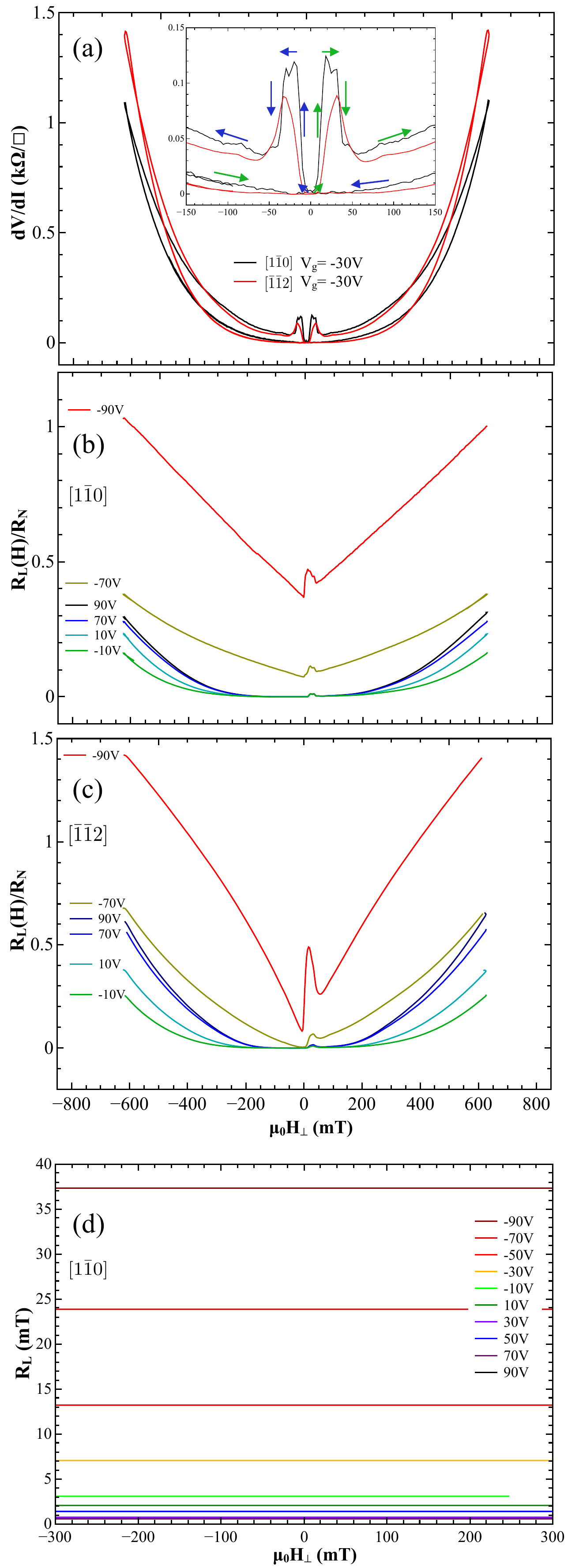}
\caption{(a)  Magnetoresistance of the [$\bar{1}\bar{1}2$] and [$1\bar{1}0$] Hall bars at 30 mK, at a back gate voltage of $V_g$=-30 V. Inset: Zoomed in plot of the MR near zero field, showing the hysteresis in magnetic field.   (b), (c) Normalized MR for the [$1\bar{1}0$] (b) and [$\bar{1}\bar{1}2$] (c) crystal directions, at 30 mK, for various gate voltages.  For clarity, only the sweeps from negative to positive magnetic field are shown. (d)  longitudinal MR along the [$1\bar{1}0$] direction }
\label{MR}
\end{figure} 

We now discuss the MR along the two crystal directions in a magnetic field perpendicular to the interface well below the superconducting transition.  Figure \ref{MR}(a) shows the MR along the two crystal directions for $V_g$=-30 V at 30 mK.  The traces for both directions are hysteretic, similar to what has been observed earlier for (001) LAO/STO devices, showing the coexistence of magnetism and superconductivity in (111) LAO/STO devices.\cite{Dikin,Mehta}  While the MR traces along the two directions appear very similar at this gate voltage, the differences are accentuated as $V_g$ is decreased.  Figs. \ref{MR}(b) and (c) show the perpendicular MR for different gate voltages along the [$1\bar{1}0$] and [$\bar{1}\bar{1}2$] crystal directions respectively.  Since the normal state resistance $R_N$ changes by three orders of magnitude over the gate voltage range, the resistance values for each gate voltage are normalized by $R_N$, which we take to be the value of $dV/dI$ at $I_{dc}=2.5$ $\mu$A at 30 mK for that value of $V_g$ (Figs. \ref{dVdI}(a) and (b)).  

Consider first the features of the MR that are similar to those observed in (001) LAO/STO devices.  The change in resistance with magnetic field is large, a significant fraction of $R_N$, particularly at negative $V_g$.  This appears similar to the (001) LAO/STO devices, where a large MR was observed in the superconducting state.\cite{Mehta} On sweeping from negative to positive magnetic field, a peak is observed in the MR at $\sim$34 mT, which grows in relative magnitude as $V_g$ is decreased.  A mirror symmetric peak is observed at $\sim$-34 mT on sweeping down in magnetic field.  Similar hysteretic peaks in the MR at small fields have been observed in the (001) LAO/STO devices,\cite{Dikin,Mehta} and have been associated with the magnetization dynamics of the ferromagnet at the interface: it is reasonable to assume that the origin of the peaks here is similar.  In contrast, at 4 K, there is almost no MR:  Fig. \ref{MR}(d) shows the MR along the [$1\bar{1}0$] direction for various gate voltages at 4 K.  The curves are flat to within our noise (MR for the [$\bar{1}\bar{1}2$] direction is similarly flat).

There are, however, some important differences between the (001) and (111) LAO/STO samples.  First, in the (001) LAO/STO devices, once a perpendicular magnetic field was applied, the samples never returned to a true zero resistance state unless they were warmed up well above the superconducting transition.\cite{Mehta}  This is consistent with the expected vortex dynamics in a two-dimensional superconductor.  In contrast, for positive values of $V_g$, the resistance of the (111) LAO/STO devices vanishes as the magnetic field is swept to zero, as can be seen from Figs. \ref{MR}(b) and (c).  Second, for larger perpendicular magnetic fields ($\geq 150$ mT), the MR in (001) LAO/STO devices for positive $V_g$ (\textit{i.e.}, in the superconducting regime) became non-hysteretic and saturated at a value corresponding to $R_N$.  As can be seen from Figs. \ref{MR}(b) and (c), the MR for the (111) devices gives no indication of saturating even over much larger field scales for either crystal direction: indeed, with $R_N$ defined as above, the resistance exceeds $R_N$ for magnetic fields greater than 600 mT in both crystal directions for $V_g=-90$ V.  Third, while the general shapes of the MR curves in perpendicular magnetic field $B$ in the (001) LAO/STO were similar in shape and curvature, changing primarily in magnitude as $V_g$ changed, the shapes of the MR curves for the (111) devices change significantly over the gate voltage range measured.  For positive $V_g$, the MR is quadratic in $B$, with positive curvature for $B>50$ mT, similar to what was observed in the (001) devices.  For large negative $V_g$, however, the curvature gradually changes; for $V_g=-90$ V, the MR for the [$1\bar{1}0$] direction is strikingly linear, while the curvature of the MR trace for the [$\bar{1}\bar{1}2$] direction becomes negative.  It is also in this gate voltage regime that the resistance at the largest fields exceeds $R_N$.  MR that is quasi-linear in $B$ has been reported recently in a number of materials with potential topological properties, where it has been suggested that it might be associated with Dirac bands in the lowest Landau level, for example see Ref. \citenum{Xiao}  The difference in our samples is that the large, quasi-linear MR is observed only at very low temperatures (it disappears at 4 K), and the quasi-linear behavior persists down to very small magnetic fields. Its appearance at low temperatures suggests it might be associated with superconductivity; on the other hand, the fact that the resistance in this gate voltage regime at even moderate magnetic fields exceeds $R_N$ indicates that its origin may be associated with normal magnetotransport that only manifests itself at the lowest temperatures.          

A fourth distinct difference between the (001) and (111) LAO/STO devices is that the MR in the (111) LAO/STO devices is different along different crystal directions, reflecting the anisotropy found in almost all the other properties.  As with those other properties, this anisotropy in the MR is most evident for negative $V_g$, as can be seen by comparing the MR curves for $V_g=-90$ V for the two crystal directions where the resistivity is higher. 

\begin{figure}[h!]
\includegraphics[width=7.5cm]{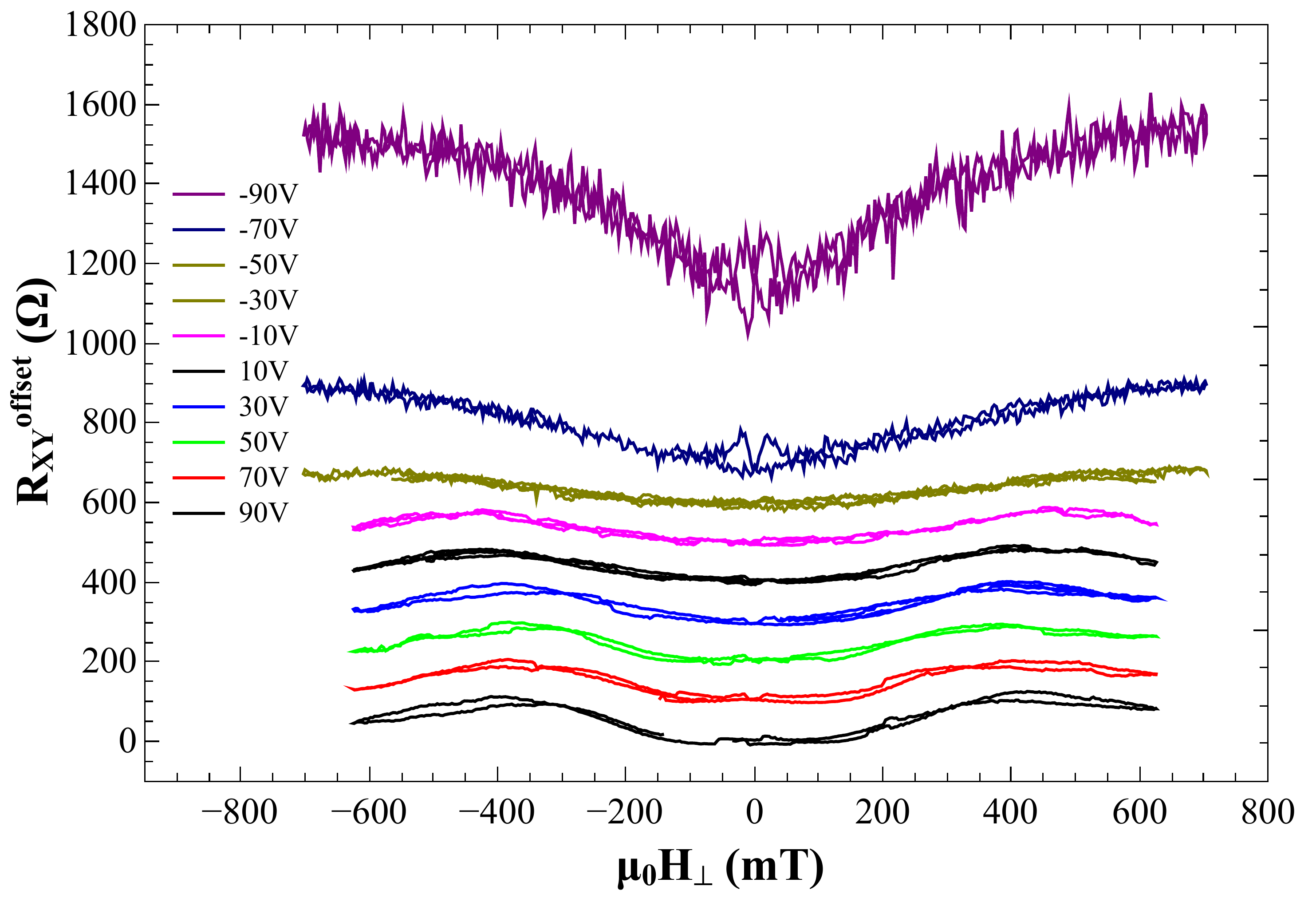}
\caption{Hall magnetoresistance for the [$1\bar{1}0$] direction at 30mK for a variety of $V_g$. For clarity each of the curves is offset from the next gate voltage by 100 $\Omega$.  }
\label{Hall}
\end{figure} 

One final difference between the (001) and (111) LAO/STO devices is seen in the Hall response in the superconducting state. Figure \ref{Hall} shows the transverse (Hall) resistance of the [$1\bar{1}0$] Hall bar for various gate voltages $V_g$ as a function of magnetic field at 30 mK.  The Hall resistance appears to be symmetric in magnetic field; in fact, there is a small antisymmetric component whose slope is approximately 1/3 of the Hall slope at 4.4 K ($\sim 35$ $\Omega$/T).  In general, a symmetric component to the transverse resistance may appear due to a small misalignment of the transverse voltage probes, which results in a fraction of the longitudinal MR appearing in the Hall measurement.  In our case, the misalignment is $\sim 1$ $\mu$m over a total length between longitudinal voltage probes of 600 $\mu$m ($\sim$0.2\%).  At $V_g$=-90 V, with this level of misalignment, we might expect to see a symmetric signal of order $\sim$ 180 $\Omega$ in the transverse Hall signal coming from the longitudinal MR, while the overall resistance change is $\sim$ 400 $\Omega$.  However, this cannot be the origin of the entire transverse MR, since the shape of the transverse MR curves do not reflect the corresponding longitudinal resistance curves, as can be seen by comparing Figs. \ref{MR}(b) and \ref{Hall}.  This is evident in all the traces, particularly the ones at large $V_g$, where the longitudinal resistance vanishes over a range of magnetic fields, and hence there should be no contribution from misaligned contacts.  It is known that the motion of vortices in superconductors can give rise to a symmetric transverse voltage, and we believe that the remaining contribution to the Hall signal is due to vortex dynamics under the influence of magnetic field and measurement current.\cite{Kopnin,Kopnin2}  We note that these data are quite different from the transverse MR observed in the superconducting state in (001) LAO/STO samples.\cite{Dikin} In that case, the transverse MR is hysteretic, as it is for large negative $V_g$ here, but the linear antisymmetric component of the transverse MR is large and comparable to its value at temperatures well above the superconducting transition. 

In conclusion, we have shown that the (111) LAO/STO interface exhibits the coexistence of ferromagnetic and superconducting phases; similar to the (001) interface this coexistence is seen most prominently in hysteretic MR.  Additionally we also observe that the $T_c$ vs $H_{\perp/\parallel}$ phase diagram is strongly hysteretic, giving further evidence of coexistence; furthermore, using these curves we have determined the thickness of the gas and $\xi_0$ are of the same order, albeit slightly smaller, than those seen in (001) LAO/STO. However, standing in stark difference to the behavior at the (001) interface, the (111) interface shows a return to a zero resistance state at low field, larger high field hysteresis, and a superconducting Hall effect that is three orders of magnitude larger than in the (001) interfaces. These differences may originate from the more disordered nature of the gas at the (111) interface, corroborated by the higher normal state resistances, or possibly from a exotic source such as different Rashba coupling the two in-plane crystal directions. \cite{Gop} More importantly the (111) interface shows strong anisotropy that is not observed in (001) interface and moreover coupled to not only the response of the system to $I_{dc}$, but also the hysteretic MR.  This coupling opens a new avenue to investigate possibly unconventional two dimensional superconductivity, and will require further work to elucidate its origin.

\begin{acknowledgments}
Work at Northwestern was funded through a grant from the U.S. Department of Energy through Grant No. DE-FG02-06ER46346.	Work at NUS was supported by the MOE Tier 1 (Grant No. R-144-000-364-112 and R-144-000-346-112) and Singapore National Research Foundation (NRF) under the Competitive Research Programs (CRP Award Nos. NRF-CRP8-2011-06, NRF-CRP10-2012-02, and NRF-CRP15-2015-01).

\end{acknowledgments}

\end{document}